\documentclass[%
 reprint,
 showpacs,preprintnumbers,
 amsmath,amssymb,
 aps,
 prb,
]{revtex4-1}
\usepackage{dcolumn}
\usepackage[dvipdfmx]{graphicx}
\usepackage{subfigure}
\usepackage{color}
\usepackage{mathrsfs}

\makeatletter
\def\btt#1{\texttt{\@backslashchar#1}}
\DeclareRobustCommand\bblash{\btt{\@backslashchar}} \makeatother

\begin{document}

\title{Time-reversal breaking topological phase without Hall electric current in a two-dimensional Dirac semimetal protected by nonsymmorphic symmetry}

\author{Tetsuro Habe}
\affiliation{Department of Physics, Osaka University, Toyonaka, Osaka 560-0043, Japan}

\date{\today}

\begin{abstract}
We investigate the topological phase derived by time-reversal breaking fields in a nonsymmorphic symmetry-protected two-dimensional Dirac semimetal.
When the nonsymmorphic symmetry is preserved even in the presence of the field, the two-dimensional electronic states change into two distinct topological phases with the insulating gap.
One phase is well-known as quantum Hall states with chiral edge modes accompanying the Hall current, but the other one is an unconventional topological phase with helical edge modes in the absence of time-reversal symmetry.
\end{abstract}


\maketitle

Two-dimensional (2D) massless Dirac fermion system in condensed matter physics has attracted much attention in a broad area of physics.
Such an excitation spectrum of electrons was first discovered in graphene, a monolayer of graphite, and the electronic states have two Dirac points, at which the cone-like conduction and valence bands, so-called Dirac cone, are touched, in the first Brillouin zone \cite{Neto2009}.
The other well-known pseudo-2D electronic states with the Dirac cone are surface semimetallic states of three-dimensional topological insulators \cite{Fu2007,Fu2007-2}.
In both cases, the gappless energy dispersion is protected by the symmetry preserved in each material: sublattice symmetry in graphene and time-reversal symmetry in surface states of the three-dimensional topological insulator \cite{Schnyder2008}.
When such a symmetry is broken by a field, the 2D gapless modes change into the topologically-nontrivial gapped electronic states \cite{Murakami2007}.
For instance, the staggered spin-orbit coupling opens an insulating gap in graphene, and leads to quantum spin Hall states, time-reversal invariant topological phase of 2D electronic system \cite{Kane2005,Kane2005-2}.
In the case of three-dimensional topological insulator, a perpendicular magnetization leads to the quantum Hall effect in the surface modes \cite{Qi2008}

Recently, Young and Kane proposed a novel 2D Dirac semimetal holding three Dirac points \cite{Young2015}.
The Dirac points are protected by time-reversal and nonsymmorphic symmetries: glide mirror symmetry or/and screw symmetry, and
appear on the boundary of the first Brillouin zone.
When the both nonsymmorphic symmetries are broken by the time-reversal invariant field, the excitation spectrum is gapped and the 2D Dirac semimetal changes into a topological or conventional insulator.

In this paper, we consider the effect of time-reversal breaking fields preserving glide mirror symmetry, a nonsymmorphic symmetry, to the electronic states in the novel 2D Dirac semimetal.
We find such a field opens an insulating gap and leads to two types of topological phases depending on the symmetric property of the field.
One topological phase is well-known as quantum Hall states characterized by the Hall current, and the other topological phase accompanies the flow of glide mirror parity: the eigenvalue of glide mirror operation, without the Hall current.
We also show the realistic fabrication for leading to the topological phase transition in the novel 2D Dirac semimetal.

\begin{figure}[htbp]
\begin{center}
 \includegraphics[width=80mm]{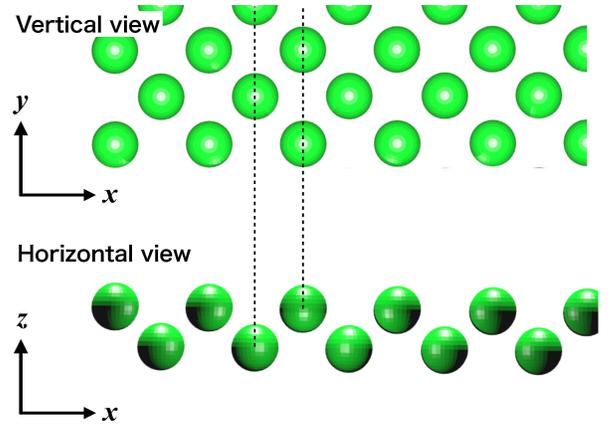}
\caption{The schematic picture of the lattice structure of the 2D Dirac semimetal protected by nonsymmorphic symmetries.
 }\label{Fig_lattice_structure}
\end{center}
\end{figure}
We consider electronic states in a 2D Dirac semimetal protected by nonsymmorphic symmetries, and investigate the effect of fields preserving glide mirror symmetry to the electronic states.
The minimal model of such a 2D material was given by Young and Kane\cite{Young2015}, and it is the square-lattice model with two sublattices displaced in the in-plane and out-of-plane directions as shown in Fig.\ \ref{Fig_lattice_structure}.
The electronic states in the semimetal are described by
\begin{align}
H_0=&2t\tau_x\cos\frac{k_x}{2}\cos\frac{k_y}{2}+t_2(\cos k_x+\cos k_y)\notag\\
&+t_{so}\tau_z(\sigma_y\sin k_x-\sigma_x\sin k_y)
,\label{tb_Hamiltonian}
\end{align}
with Pauli matrices $\sigma$ and $\tau$ for the spin and the sublattice, respectively.
Here, $t$ and $t_2$ are nearest and second-nearest neighbor hopping matrices, respectively, and $t_{so}$ is the coupling constant of the spin-orbit coupling.
The electronic states are doubly degenerated at each wave vector because of glide mirror symmetry $\sigma_z\tau_x$, and have three Dirac points, 
where the energy dispersion is a linear function of the relative wave vector $\boldsymbol{p}$ with respect to the point.
Thus, the low-energy electronic states can be described by the $2\times2$ effective Hamiltonian for each glide mirror parity $\xi_z=\sigma_z\tau_x$ with $s_z=\tau_x$ around the Dirac points\cite{Habe2017}, and that around $X_1=(-\pi,0)$ and $X_2=(0,-\pi)$, called valleys, can be represented by
\begin{align}
H_\zeta(\boldsymbol{p})=s_z(u_\zeta p_x+u'_\zeta p_y)-\zeta v_{so}(\xi_zs_yp_x+s_xp_y),\label{Effective_Hamiltonian}
\end{align}
with the valley dependent parameters $u_\zeta$ and $u_\zeta'$ for $\zeta=1$ at $X_1=(-\pi,0)$ and $\zeta=-1$ at $X_2=(0,-\pi)$.
The effective Hamiltonian around the other Dirac point $M=(\pi,\pi)$ can be obtained by $u_\zeta=u_\zeta'=0$ and $\xi_z\rightarrow-\xi_z$.
Therefore, we can analyze the effect of a homogeneous field coupling to the electronic states, excepting the electron-electron interaction, around each Dirac point by using the general form of Eq.(\ref{Effective_Hamiltonian}).

The uniform fields preserving glide mirror symmetry can be represented by the diagonal operator in the $\xi$ space, and they do not open a gap at the Dirac points as long as time-reversal symmetry $\mathcal{T}=i\tau_z\sigma_y\mathcal{K}$ remains\cite{Young2015}.
For instance, the simplest time-reversal invariant field, which is proportional to identity, describes the shift of the Fermi energy $\varepsilon_FI_{4\times4}$ i.e., the change in the charge density, but it does not change the energy dispersion and the electronic states.
The shift of the Fermi energy and the split of glide mirror parity $U\xi_z=U\tau_x\sigma_z$, both fields preserve time-reversal symmetry, leave the gapless energy dispersion, i.e., they can not induce topological insulating phase, and the later field leads to the imbalance of $\xi_z=\pm1$ in the electric flow induced by an electric field because of the non-equal Fermi surface for $\xi_z=\pm1$.

Any time-reversal symmetry-breaking field preserving glide mirror symmetry is described by the Zeeman-like field $-\boldsymbol{M}(\xi_z,\zeta)\cdot\boldsymbol{s}$ in the $s$ space, and we show that the effect to electronic states can be evaluated by the Hamiltonian Eq.(\ref{Effective_Hamiltonian}) plus  $-M_zs_z$.
In this representation, time-reversal operator is given by $\mathcal{T}=i\xi_z s_y\mathcal{K}$ with complex conjugation $\mathcal{K}$, and thus the time-reversal symmetry-breaking field can be written by a spinful operator in the $s$ space. 
In the ordinary two-dimensional Dirac fermion system, the in-plane field, in general, can be eliminated in the Hamiltonian apparently by a shift of the Dirac point\cite{Yokoyama2010,Habe2012} but the in-plane component plays a similar role of the out-of-plane component in the 2D Dirac semimetal.
The  Zeeman-like field can be rewritten by the $z$ component with a shift of the Dirac point in general,
\begin{align}
H_m'=-s_z\left(M_z+\zeta\frac{u_\zeta'}{v_{so}}M_x+\zeta\xi_z\frac{u_\zeta}{v_{so}}M_y\right),\label{general_form_of_field}
\end{align}
with a translation of the wave number$q_x=p_x+\zeta\xi_z(M_y/v_{so})$ and $q_y=p_y+\zeta(M_x/v_{so})$.
At the $M$ valley, however the in-plane field only shift the Dirac point because of $u_\zeta=u_\zeta'=0$.

The time-reversal symmetry-breaking field opens an energy gap in the electronic excitation spectrum and the gap is depending on the field strength and the ratio of the hopping matrix and the spin-orbit coupling.
The Zeeman-like field in any direction, the generalized form of time-reversal symmetry-breaking field, can be rewritten by the out-of-plane field $-M_z(\xi_z,\zeta)s_z$ in Eq.\ (\ref{general_form_of_field}), and the electronic states, in general, have an insulating gap in the excitation spectrum where the energy dispersion is particle-hole symmetric $\varepsilon=\pm d_{\xi_z,\boldsymbol{p}}$ and given by
\begin{align}
d_{\xi_z,\boldsymbol{p}}=\sqrt{[\bar{u}p\cos(\theta-\theta_\zeta)-M_z(\xi_z,\zeta)]^2+v_{so}^2p^2}
\end{align}
with $\cos\theta_\zeta=u_\zeta/\bar{u}$ and $\bar{u}=\sqrt{{u_\zeta}^2+{u_\zeta'}^2}$.
Here, the energy gap is proportional to the strength of the Zeeman field,
\begin{align}
\Delta\varepsilon=\frac{2\left|M_z(\xi_z,\zeta)\right|}{\sqrt{\bar{u}^2/v_{so}^2+1}},\label{Eq_band_gap}
\end{align}
and the top of the dispersion shifts to
\begin{align}
\boldsymbol{p}=\frac{\bar{u}}{\bar{u}^2+v_{so}^2}M_z(\xi_z,\zeta)(\cos\theta_\zeta,\sin\theta_\zeta).
\end{align}
In the strong limit of the spin-orbit coupling $\bar{u}/v_{so}\ll1$, the energy dispersion is asymptotically close to that of the ordinary 2D Dirac fermion\cite{Murakami2007}.

The topological property of the two-dimensional insulating phase is characterized by the Chern number.
The Chern number is defined by the integral of the Berry curvature $\Omega(\boldsymbol{p})=i[\nabla_{\boldsymbol{p}}\times\langle u(\boldsymbol{p})|\nabla_{\boldsymbol{p}}|u(\boldsymbol{p})\rangle]_z$ for the electronic state $|u(\boldsymbol{p})\rangle$ in the occupied band\cite{Haldane2004,Qi2006}, i.e., the valence band,
$C=\frac{1}{2\pi}\int_{\mathrm{VB}}{d^2\boldsymbol{p}\;}\Omega(\boldsymbol{p})$,
where VB means the valence band, and it is equivalent to the quantized Hall conductivity,
\begin{align}
\sigma_{xy}^e=\frac{e^2}{h}C.
\end{align}
We calculate the quantized Hall conductivity to characterize the electric states.
The Hall conductivity also can be obtained as a sum of those for Dirac electrons described by Eq.\ (\ref{Effective_Hamiltonian}) plus the Zeeman-like field at every Dirac points, where they are equivalent to the asymptotic value of the Berry phase close to the valence band edge as shown in Appendix \ref{Ap1}.
The Berry phase can be calculated from the phase of the wave function
\begin{align}
\phi(\varepsilon_F)=\oint_{C_F}d\boldsymbol{\ell}_{\boldsymbol{p}}\cdot\langle u(\boldsymbol{p})|\nabla_{\boldsymbol{p}}|u(\boldsymbol{p})\rangle,
\end{align}
in the Fermi surface $C_F$\cite{Haldane2004}.
When the Fermi energy $\varepsilon_F$ is close to the valence band edge, we can describe the electronic states in $\varepsilon_F$ by Eq.\ (\ref{Effective_Hamiltonian}) and thus we can obtain the geometrical phase in each $(\zeta,\xi_z)$
\begin{align}
\phi_{\xi_z,\zeta}=\pi\frac{\xi_zM_z(\xi_z,\zeta)}{\varepsilon_F\sqrt{\bar{u}^2/v_{so}^2+1}},
\end{align}
with the Fermi energy $\varepsilon_F$ with respect to the center of the gap.
The asymptotic value $\phi_{\xi_z,\zeta}^{0}$ of the geometrical phase can be obtained by $\varepsilon_F\rightarrow-\Delta\varepsilon/2$ in Eq.\ (\ref{Eq_band_gap}).

We first discuss the topological characteristic of the gapped states by the quantized Hall conductivity around each symmetrical point. The conductivity is given by the summation of the geometrical phase $\phi^0_{\xi_z,\zeta}$ for $\xi_z$,
\begin{align}
\sigma_{xy}^e(\zeta)=\frac{e^2}{2\pi h}\sum_{\xi_z}\phi^0_{\zeta,\xi_z}.\label{Eq_Hall_conductivity_e}
\end{align}
Thus, if the time-reversal symmetry breaking field is asymmetric about $\xi_z$, the Hall conductivity becomes $\pm e^2/h$ at each symmetrical point.
The non-zero Hall conductivity is attributed to the doubly degenerated chiral edge modes corresponding to the same sign geometrical phase $\phi^0_{\xi_z,\zeta}$ for $\xi_z=\pm1$.

When the time-reversal breaking field is symmetric about $\xi_z$, the Hall conductivity is absent in each valley but we can show that the gapped states can be characterized by a topological quantity defined by
\begin{align}
c^\xi(\zeta)=\frac{1}{2\pi}\sum_{\xi_z}\xi_z\phi^0_{\zeta,\xi_z}.
\end{align}
The non-zero topological number means that the electronic states have the opposite chiral edge modes in two $\xi_z$, i.e., the topological phase has the helical edge mode.
The absence of the Hall conductivity is attributed to the helical edge states, but we can define the virtual Hall conductivity of the glide mirror parity $\xi_z$ as
\begin{align}
\sigma_{xy}^\xi(\zeta)=\frac{e^2}{h}c^\xi(\zeta).\label{Eq_Hall_conductivity_xi}
\end{align}

\begin{table}[htb]
\caption{The relation between the symmetrical characteristics of a field with $\sigma_\mu\tau_\nu$ and the induced Hall conductivity. The absence of the symmetry is indicated by 0 and the presence is represented by 1. The box for the Hall conductivity is filled by 1 for a quantized conductivity in the unit of $e^2/h$ and 0 for absence.}
\begin{center}
\begin{tabular}{c c c c c c }
\hline \hline
&$\mathcal{T}$&$\mathcal{T}_2$&$\sigma_{xy}^e$&$\sigma_{xy}^\xi$\\ \hline
$\sigma_z,\;\sigma_x\tau_z,\;\sigma_y\tau_z$&0&0&1&0\\  
$\tau_x,\;\sigma_x\tau_y,\;\sigma_y\tau_y$&0&1&0&1 \\ 
$\sigma_z\tau_x$&1&0&0&0\\ \hline \hline
\end{tabular}\label{tab_symmetry_Hall}
\end{center}
\end{table}
We briefly summarize the relation between the topologically non-trivial states around each symmetrical point and the symmetrical characteristics of the homogeneous field represented by $U\sigma_\mu\tau_\nu$.
The glide mirror symmetry-conserving fields can be classified by time-reversal symmetry $\mathcal{T}=i\tau_z\sigma_y\mathcal{K}$ and pseudo time-reversal $\mathcal{T}_2=\mathcal{T}\tau_z$, and we evaluate the topological property by the two types of Hall conductivities as shown in Table \ref{tab_symmetry_Hall}. 
The time-reversal symmetry-breaking field $U\sigma_\mu\tau_\nu$, in general, gives a topological characteristic to the electronic states, and  such gapped electronic states can be characterized by $\sigma_{xy}^e$ for $\mathcal{T}_2$-breaking field and by $\sigma_{xy}^\xi$ for $\mathcal{T}_2$-conserving field.

\begin{figure}[htbp]
\begin{center}
 \includegraphics[width=70mm]{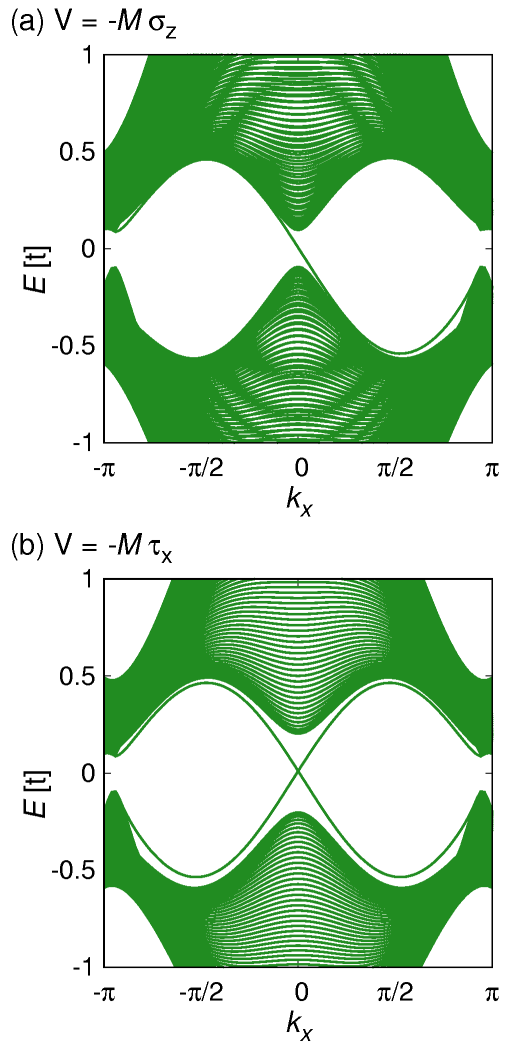}
\caption{The numerically calculated band structure in the nonsymmorphic symmetry-protected 2D Dirac semimetal with the hard wall boundary along the $x$ axis. The band structure consists of the energy dispersion for the extended states and the edge states in one side. The band gap is induced by $V=-M\sigma_z$ in (a) and $V=-M\tau_x$ in (b) with $M=0.2t$ with the parameters of $t_2=0.1t$ and $t_{so}=0.5t$. 
 }\label{Fig_Band}
\end{center}
\end{figure}
Next, we discuss the global topological characteristic of the electronic states in the presence of time-reversal breaking field, and investigate the relation between the topological phase and the number of the Dirac points.
Such a global property can be evaluated by summing $\sigma_{xy}^\alpha(\zeta)$ in all symmetrical points $X_1$, $X_2$, and $M$, and thus the Hall conductivity remains in non-zero integer because of odd number of Dirac points.
We show the band structure of such topological states induced by $\mathcal{T}_2$-breaking and $\mathcal{T}_2$-conserving fields in Figs.\ \ref{Fig_Band} (a) and (b), respectively.
In both cases, we calculate the band structure with the hard wall boundary along the $x$ axis, and the chiral and helical edge modes can be found corresponding to the symmetry of field.
Moreover, the edge mode is absent around $k_x=\pm \pi$ because the electronic states around $M=(\pi,\pi)$ and $X_1=(\pi,0)$ have opposite topological numbers.

Finally, we discuss the realistic fabrication to introduce such fields for realizing the two time-reversal breaking topological phases.
The conventional Hall states in Fig.\ \ref{Fig_Band} (a) can be realized by a magnetic proximity effect from an ferromagnetic slab attached to the semimetal.
The out-of-plane Zeeman field can be represented by $U\sigma_z$ and classified into the field leading to the quantized Hall conductivity $\sigma_{xy}^e$ as shown in Table\ \ref{tab_symmetry_Hall}.
The staggered Zeeman field $\sigma_\mu\tau_z$ for $\mu=x,y$, the other field inducing non-zero $\sigma^e_{xy}$, can be introduced in the sandwich structure of two ferromagnets with the opposite in-plane magnetization due to the displacement of the sublattices in the out-of-plane direction as shown in Fig.\ \ref{Fig_lattice_structure}.

The field represented by $U\tau_x$, which is the occasion of the helical edge modes, can be introduced by applying an in-plane magnetic field in the $X_j$ and $M$ valleys.
The in-plane magnetic field can be represented by an in-plane gauge field depending on the out-of-plane position, and thus the electron affected by the difference of the gauge field in the hopping process between the sublattices which are displaced in the out-of-plane direction.
When the magnetic field is given by $(B_x,B_y,0)$, the gauge field can be represented by $\boldsymbol{A}=(B_yz,-B_xz,0)$ and the inter-sublattice hopping matrix proportional to $\tau_x$ changes into
\begin{align}
u_\zeta p_x+u_\zeta' p_y\rightarrow u_\zeta\left(p_x+\frac{ed_z^2}{2\hbar a}B_y\right)+u_\zeta'\left(p_y-\frac{ed_z^2}{2\hbar a}B_x\right),
\end{align}
with the lattice constant $a$ and the distance $d_z$ between the sublattices in Fig.\ \ref{Fig_lattice_structure}.
At $M$ point, we have the other way to induce such a field by the lattice deformation\cite{Chen2014,Young2015,Liu2016}.

In the conclusion, we investigate the topological phase in the nonsymmorphic symmetry-protected 2D Dirac semimetal in the presence of time-reversal breaking field.
We find two distinct topological phases corresponding to the symmetrical characteristics of the field about pseudo time-reversal symmetry $\mathcal{T}_2$.
One is the conventional quantum Hall states, and the other is the unconventional time-reversal breaking topological phase with helical edge modes.

\appendix
\section{Chern number and electronic states around symmetrical momenta}\label{Ap1}
In this section, we show that the Chern number in the 2D Dirac semimetal can be evaluated by analyzing the electric states around the Dirac points $M$, $X_1$, and $X_2$ described by the Hamiltonian in Eq.\ (\ref{Effective_Hamiltonian}).

The Chern number is a robust topological number, which does not require specific symmetries, and thus it is unchanged under the continuous deformation of Hamiltonian without gap closing.
Thus, we consider a wave vector-dependent exchange potential $M_{\mu}(\boldsymbol{k})=f_{\epsilon,\delta}(\boldsymbol{k})$, instead of the homogeneous exchange potential, which is non-zero value around the symmetrical points $\boldsymbol{k}_j$ for $M$, $X_1$, and $X_2$ as
\begin{align}
f_{\epsilon,\delta}(\boldsymbol{k})=\begin{cases}
M_0&(|\boldsymbol{k}-\boldsymbol{k}_j|<\epsilon)\\
0&(\epsilon+\delta<|\boldsymbol{k}-\boldsymbol{k}_j|)
\end{cases},
\end{align}
with parameters $\epsilon$ and $\delta$, where $f_{\epsilon,\delta}$ is smoothly varying between $M_0$ and $0$ in the region $\epsilon\leq|\boldsymbol{k}-\boldsymbol{k}_j|\leq\epsilon+\delta$.
Since the original Hamiltonian plus this local potential is continuously connecting to that plus a homogenous exchange potential by $\delta\rightarrow\infty$, the Chern number is unchanged between two models of exchange potential.

If we assume that electronic states in the valence bands are unoccupied in $S_{q}$ where $|\boldsymbol{k}-\boldsymbol{k}_j|<q$, the Berry phase is given by the difference of the Chern number $C$ of the valence band and the integral of the Berry curvature $\theta(q)/2\pi$ in $S_{q}$.
Since $C-\theta(q)/2\pi$ can be obtained by integral of Berry curvature of electronic states out of $S_q$, it is exactly zero as long as $\delta+\epsilon<q$ because the exchange field is absent out of $S_q$ and the electronic states are equivalent to those in time-reversal invariant system.
Therefore, the Chern number is equivalent to the sum of integral of Berry curvature in the vicinities of symmetrical points $\boldsymbol{k}_j$.

The Hamiltonian describing electric states in $S_q$ asymptotically reaches Eq.\ (\ref{Effective_Hamiltonian}) by reducing $\epsilon$ and $\delta$, where the area with non-zero exchange field can be decreased as long as $0<\epsilon<\delta$ without closing the band gap.
Thus, we can evaluate the Chern number by electronic states described by Eq.\ (\ref{Effective_Hamiltonian}) with the exchange field around every symmetrical points $\boldsymbol{k}_j$. 
Moreover, we can connect the electric structure described by Eq.\ (\ref{Effective_Hamiltonian}), the time-reversal invariant Hamiltonian, out of $S_q$ and then expand the area with non-zero Zeeman-like field in this additional area. 
Here, the decay in $M_\mu(\boldsymbol{k})$ is naturally satisfied on electric states in the limit of $|\boldsymbol{k}-\boldsymbol{k}_j|\rightarrow\infty$ because the exchange field dependence of electric states is given by $M_\mu(\boldsymbol{k})/\delta k$ for a large $\delta k\equiv|\boldsymbol{k}-\boldsymbol{k}_j|$.

Therefore, we can evaluate the Chen number by summing the Berry phases of electronic states described by Eq.\ (\ref{Effective_Hamiltonian}) plus a Zeeman-like field for every symmetrical points.

\bibliography{2D_Dirac}

\providecommand{\noopsort}[1]{}\providecommand{\singleletter}[1]{#1}%
\begin{thebibliography}{16}%
\makeatletter
\providecommand \@ifxundefined [1]{%
 \@ifx{#1\undefined}
}%
\providecommand \@ifnum [1]{%
 \ifnum #1\expandafter \@firstoftwo
 \else \expandafter \@secondoftwo
 \fi
}%
\providecommand \@ifx [1]{%
 \ifx #1\expandafter \@firstoftwo
 \else \expandafter \@secondoftwo
 \fi
}%
\providecommand \natexlab [1]{#1}%
\providecommand \enquote  [1]{``#1''}%
\providecommand \bibnamefont  [1]{#1}%
\providecommand \bibfnamefont [1]{#1}%
\providecommand \citenamefont [1]{#1}%
\providecommand \href@noop [0]{\@secondoftwo}%
\providecommand \href [0]{\begingroup \@sanitize@url \@href}%
\providecommand \@href[1]{\@@startlink{#1}\@@href}%
\providecommand \@@href[1]{\endgroup#1\@@endlink}%
\providecommand \@sanitize@url [0]{\catcode `\\12\catcode `\$12\catcode
  `\&12\catcode `\#12\catcode `\^12\catcode `\_12\catcode `\%12\relax}%
\providecommand \@@startlink[1]{}%
\providecommand \@@endlink[0]{}%
\providecommand \url  [0]{\begingroup\@sanitize@url \@url }%
\providecommand \@url [1]{\endgroup\@href {#1}{\urlprefix }}%
\providecommand \urlprefix  [0]{URL }%
\providecommand \Eprint [0]{\href }%
\providecommand \doibase [0]{http://dx.doi.org/}%
\providecommand \selectlanguage [0]{\@gobble}%
\providecommand \bibinfo  [0]{\@secondoftwo}%
\providecommand \bibfield  [0]{\@secondoftwo}%
\providecommand \translation [1]{[#1]}%
\providecommand \BibitemOpen [0]{}%
\providecommand \bibitemStop [0]{}%
\providecommand \bibitemNoStop [0]{.\EOS\space}%
\providecommand \EOS [0]{\spacefactor3000\relax}%
\providecommand \BibitemShut  [1]{\csname bibitem#1\endcsname}%
\let\auto@bib@innerbib\@empty
\bibitem [{\citenamefont {Castro~Neto}\ \emph {et~al.}(2009)\citenamefont
  {Castro~Neto}, \citenamefont {Guinea}, \citenamefont {Peres}, \citenamefont
  {Novoselov},\ and\ \citenamefont {Geim}}]{Neto2009}%
  \BibitemOpen
  \bibfield  {author} {\bibinfo {author} {\bibfnamefont {A.~H.}\ \bibnamefont
  {Castro~Neto}}, \bibinfo {author} {\bibfnamefont {F.}~\bibnamefont {Guinea}},
  \bibinfo {author} {\bibfnamefont {N.~M.~R.}\ \bibnamefont {Peres}}, \bibinfo
  {author} {\bibfnamefont {K.~S.}\ \bibnamefont {Novoselov}}, \ and\ \bibinfo
  {author} {\bibfnamefont {A.~K.}\ \bibnamefont {Geim}},\ }\href@noop {}
  {\bibfield  {journal} {\bibinfo  {journal} {Rev. Mod. Phys.}\ }\textbf
  {\bibinfo {volume} {81}},\ \bibinfo {pages} {109} (\bibinfo {year}
  {2009})}\BibitemShut {NoStop}%
\bibitem [{\citenamefont {Fu}\ \emph {et~al.}(2007)\citenamefont {Fu},
  \citenamefont {Kane},\ and\ \citenamefont {Mele}}]{Fu2007}%
  \BibitemOpen
  \bibfield  {author} {\bibinfo {author} {\bibfnamefont {L.}~\bibnamefont
  {Fu}}, \bibinfo {author} {\bibfnamefont {C.~L.}\ \bibnamefont {Kane}}, \ and\
  \bibinfo {author} {\bibfnamefont {E.~J.}\ \bibnamefont {Mele}},\ }\href@noop
  {} {\bibfield  {journal} {\bibinfo  {journal} {Phys.\ Rev. \ Lett.}\ }\textbf
  {\bibinfo {volume} {98}},\ \bibinfo {pages} {106803} (\bibinfo {year}
  {2007})}\BibitemShut {NoStop}%
\bibitem [{\citenamefont {Fu}\ and\ \citenamefont {Kane}(2007)}]{Fu2007-2}%
  \BibitemOpen
  \bibfield  {author} {\bibinfo {author} {\bibfnamefont {L.}~\bibnamefont
  {Fu}}\ and\ \bibinfo {author} {\bibfnamefont {C.~L.}\ \bibnamefont {Kane}},\
  }\href@noop {} {\bibfield  {journal} {\bibinfo  {journal} {Phys.\ Rev. \ B.}\
  }\textbf {\bibinfo {volume} {76}},\ \bibinfo {pages} {045302} (\bibinfo
  {year} {2007})}\BibitemShut {NoStop}%
\bibitem [{\citenamefont {Schnyder}\ \emph {et~al.}(2008)\citenamefont
  {Schnyder}, \citenamefont {Ryu}, \citenamefont {Furusaki},\ and\
  \citenamefont {Ludwig}}]{Schnyder2008}%
  \BibitemOpen
  \bibfield  {author} {\bibinfo {author} {\bibfnamefont {A.~P.}\ \bibnamefont
  {Schnyder}}, \bibinfo {author} {\bibfnamefont {S.}~\bibnamefont {Ryu}},
  \bibinfo {author} {\bibfnamefont {A.}~\bibnamefont {Furusaki}}, \ and\
  \bibinfo {author} {\bibfnamefont {A.~W.~W.}\ \bibnamefont {Ludwig}},\
  }\href@noop {} {\bibfield  {journal} {\bibinfo  {journal} {Phys. \ Rev. \ B}\
  }\textbf {\bibinfo {volume} {78}},\ \bibinfo {pages} {195125} (\bibinfo
  {year} {2008})}\BibitemShut {NoStop}%
\bibitem [{\citenamefont {Murakami}(2007)}]{Murakami2007}%
  \BibitemOpen
  \bibfield  {author} {\bibinfo {author} {\bibfnamefont {S.}~\bibnamefont
  {Murakami}},\ }\href@noop {} {\bibfield  {journal} {\bibinfo  {journal} {New.
  J. Phys.}\ }\textbf {\bibinfo {volume} {9}},\ \bibinfo {pages} {356}
  (\bibinfo {year} {2007})}\BibitemShut {NoStop}%
\bibitem [{\citenamefont {Kane}\ and\ \citenamefont
  {Mele}(2005{\natexlab{a}})}]{Kane2005}%
  \BibitemOpen
  \bibfield  {author} {\bibinfo {author} {\bibfnamefont {C.~L.}\ \bibnamefont
  {Kane}}\ and\ \bibinfo {author} {\bibfnamefont {E.~J.}\ \bibnamefont
  {Mele}},\ }\href {\doibase 10.1103/PhysRevLett.95.226801} {\bibfield
  {journal} {\bibinfo  {journal} {Phys. Rev. Lett.}\ }\textbf {\bibinfo
  {volume} {95}},\ \bibinfo {pages} {226801} (\bibinfo {year}
  {2005}{\natexlab{a}})}\BibitemShut {NoStop}%
\bibitem [{\citenamefont {Kane}\ and\ \citenamefont
  {Mele}(2005{\natexlab{b}})}]{Kane2005-2}%
  \BibitemOpen
  \bibfield  {author} {\bibinfo {author} {\bibfnamefont {C.~L.}\ \bibnamefont
  {Kane}}\ and\ \bibinfo {author} {\bibfnamefont {E.~J.}\ \bibnamefont
  {Mele}},\ }\href {\doibase 10.1103/PhysRevLett.95.146802} {\bibfield
  {journal} {\bibinfo  {journal} {Phys. Rev. Lett.}\ }\textbf {\bibinfo
  {volume} {95}},\ \bibinfo {pages} {146802} (\bibinfo {year}
  {2005}{\natexlab{b}})}\BibitemShut {NoStop}%
\bibitem [{\citenamefont {Qi}\ \emph {et~al.}(2008)\citenamefont {Qi},
  \citenamefont {Hughes},\ and\ \citenamefont {Zhang}}]{Qi2008}%
  \BibitemOpen
  \bibfield  {author} {\bibinfo {author} {\bibfnamefont {X.-L.}\ \bibnamefont
  {Qi}}, \bibinfo {author} {\bibfnamefont {T.~L.}\ \bibnamefont {Hughes}}, \
  and\ \bibinfo {author} {\bibfnamefont {S.-C.}\ \bibnamefont {Zhang}},\
  }\href@noop {} {\bibfield  {journal} {\bibinfo  {journal} {Phys. \ Rev.\ B.}\
  }\textbf {\bibinfo {volume} {78}},\ \bibinfo {pages} {195424} (\bibinfo
  {year} {2008})}\BibitemShut {NoStop}%
\bibitem [{\citenamefont {Young}\ and\ \citenamefont {Kane}(2015)}]{Young2015}%
  \BibitemOpen
  \bibfield  {author} {\bibinfo {author} {\bibfnamefont {S.~M.}\ \bibnamefont
  {Young}}\ and\ \bibinfo {author} {\bibfnamefont {C.~L.}\ \bibnamefont
  {Kane}},\ }\href {\doibase 10.1103/PhysRevLett.115.126803} {\bibfield
  {journal} {\bibinfo  {journal} {Phys. Rev. Lett.}\ }\textbf {\bibinfo
  {volume} {115}},\ \bibinfo {pages} {126803} (\bibinfo {year}
  {2015})}\BibitemShut {NoStop}%
\bibitem [{\citenamefont {Habe}(2017)}]{Habe2017}%
  \BibitemOpen
  \bibfield  {author} {\bibinfo {author} {\bibfnamefont {T.}~\bibnamefont
  {Habe}},\ }\href {\doibase 10.1103/PhysRevB.95.115405} {\bibfield  {journal}
  {\bibinfo  {journal} {Phys. Rev. B}\ }\textbf {\bibinfo {volume} {95}},\
  \bibinfo {pages} {115405} (\bibinfo {year} {2017})}\BibitemShut {NoStop}%
\bibitem [{\citenamefont {Yokoyama}\ \emph {et~al.}(2010)\citenamefont
  {Yokoyama}, \citenamefont {Tanaka},\ and\ \citenamefont
  {Nagaosa}}]{Yokoyama2010}%
  \BibitemOpen
  \bibfield  {author} {\bibinfo {author} {\bibfnamefont {T.}~\bibnamefont
  {Yokoyama}}, \bibinfo {author} {\bibfnamefont {Y.}~\bibnamefont {Tanaka}}, \
  and\ \bibinfo {author} {\bibfnamefont {N.}~\bibnamefont {Nagaosa}},\
  }\href@noop {} {\bibfield  {journal} {\bibinfo  {journal} {Phys. \ Rev. \ B}\
  }\textbf {\bibinfo {volume} {81}},\ \bibinfo {pages} {121401(R)} (\bibinfo
  {year} {2010})}\BibitemShut {NoStop}%
\bibitem [{\citenamefont {Habe}\ and\ \citenamefont {Asano}(2012)}]{Habe2012}%
  \BibitemOpen
  \bibfield  {author} {\bibinfo {author} {\bibfnamefont {T.}~\bibnamefont
  {Habe}}\ and\ \bibinfo {author} {\bibfnamefont {Y.}~\bibnamefont {Asano}},\
  }\href {\doibase 10.1103/PhysRevB.85.195325} {\bibfield  {journal} {\bibinfo
  {journal} {Phys. Rev. B}\ }\textbf {\bibinfo {volume} {85}},\ \bibinfo
  {pages} {195325} (\bibinfo {year} {2012})}\BibitemShut {NoStop}%
\bibitem [{\citenamefont {Haldane}(2004)}]{Haldane2004}%
  \BibitemOpen
  \bibfield  {author} {\bibinfo {author} {\bibfnamefont {F.~D.~M.}\
  \bibnamefont {Haldane}},\ }\href {\doibase 10.1103/PhysRevLett.93.206602}
  {\bibfield  {journal} {\bibinfo  {journal} {Phys. Rev. Lett.}\ }\textbf
  {\bibinfo {volume} {93}},\ \bibinfo {pages} {206602} (\bibinfo {year}
  {2004})}\BibitemShut {NoStop}%
\bibitem [{\citenamefont {Qi}\ \emph {et~al.}(2006)\citenamefont {Qi},
  \citenamefont {Wu},\ and\ \citenamefont {Zhang}}]{Qi2006}%
  \BibitemOpen
  \bibfield  {author} {\bibinfo {author} {\bibfnamefont {X.-L.}\ \bibnamefont
  {Qi}}, \bibinfo {author} {\bibfnamefont {Y.-S.}\ \bibnamefont {Wu}}, \ and\
  \bibinfo {author} {\bibfnamefont {S.-C.}\ \bibnamefont {Zhang}},\ }\href
  {\doibase 10.1103/PhysRevB.74.085308} {\bibfield  {journal} {\bibinfo
  {journal} {Phys. Rev. B}\ }\textbf {\bibinfo {volume} {74}},\ \bibinfo
  {pages} {085308} (\bibinfo {year} {2006})}\BibitemShut {NoStop}%
\bibitem [{\citenamefont {Chen}\ and\ \citenamefont {Kee}(2014)}]{Chen2014}%
  \BibitemOpen
  \bibfield  {author} {\bibinfo {author} {\bibfnamefont {Y.}~\bibnamefont
  {Chen}}\ and\ \bibinfo {author} {\bibfnamefont {H.-Y.}\ \bibnamefont {Kee}},\
  }\href {\doibase 10.1103/PhysRevB.90.195145} {\bibfield  {journal} {\bibinfo
  {journal} {Phys. Rev. B}\ }\textbf {\bibinfo {volume} {90}},\ \bibinfo
  {pages} {195145} (\bibinfo {year} {2014})}\BibitemShut {NoStop}%
\bibitem [{\citenamefont {Liu}\ \emph {et~al.}(2016)\citenamefont {Liu},
  \citenamefont {Kriegner}, \citenamefont {Horak}, \citenamefont {Puggioni},
  \citenamefont {Rayan~Serrao}, \citenamefont {Chen}, \citenamefont {Yi},
  \citenamefont {Frontera}, \citenamefont {Holy}, \citenamefont {Vishwanath},
  \citenamefont {Rondinelli}, \citenamefont {Marti},\ and\ \citenamefont
  {Ramesh}}]{Liu2016}%
  \BibitemOpen
  \bibfield  {author} {\bibinfo {author} {\bibfnamefont {J.}~\bibnamefont
  {Liu}}, \bibinfo {author} {\bibfnamefont {D.}~\bibnamefont {Kriegner}},
  \bibinfo {author} {\bibfnamefont {L.}~\bibnamefont {Horak}}, \bibinfo
  {author} {\bibfnamefont {D.}~\bibnamefont {Puggioni}}, \bibinfo {author}
  {\bibfnamefont {C.}~\bibnamefont {Rayan~Serrao}}, \bibinfo {author}
  {\bibfnamefont {R.}~\bibnamefont {Chen}}, \bibinfo {author} {\bibfnamefont
  {D.}~\bibnamefont {Yi}}, \bibinfo {author} {\bibfnamefont {C.}~\bibnamefont
  {Frontera}}, \bibinfo {author} {\bibfnamefont {V.}~\bibnamefont {Holy}},
  \bibinfo {author} {\bibfnamefont {A.}~\bibnamefont {Vishwanath}}, \bibinfo
  {author} {\bibfnamefont {J.~M.}\ \bibnamefont {Rondinelli}}, \bibinfo
  {author} {\bibfnamefont {X.}~\bibnamefont {Marti}}, \ and\ \bibinfo {author}
  {\bibfnamefont {R.}~\bibnamefont {Ramesh}},\ }\href {\doibase
  10.1103/PhysRevB.93.085118} {\bibfield  {journal} {\bibinfo  {journal} {Phys.
  Rev. B}\ }\textbf {\bibinfo {volume} {93}},\ \bibinfo {pages} {085118}
  (\bibinfo {year} {2016})}\BibitemShut {NoStop}%
\end{thebibliography}%

\end{document}